\documentclass[12pt]{article}
\usepackage[utf8]{inputenc}
\usepackage{amssymb}
\usepackage{amsthm}
\usepackage{amsmath}
\usepackage{booktabs}
\usepackage{fancyhdr}
\usepackage{graphicx}
\usepackage{natbib}
\usepackage{mathtools}
\usepackage{breqn}
\usepackage{dsfont}
\usepackage{adjustbox}
\usepackage{multirow}
\usepackage{bbold}
\usepackage[margin=1.25in]{geometry}
\usepackage{setspace}
\RequirePackage[colorlinks,citecolor=blue,linkcolor=blue,urlcolor=blue,pagebackref]{hyperref}

\theoremstyle{plain}
\newtheorem{theorem}{Theorem}
\newtheorem{lemma}[theorem]{Lemma}
\newtheorem{proposition}[theorem]{Proposition}

\theoremstyle{definition}
\newtheorem{definition}{Definition}

\newtheorem{assumption}[definition]{Assumption}

\theoremstyle{definition}
\newtheorem{example}{Example}

\newtheorem{remark}{Remark}

\title{Peer effect analysis with latent processes}
\author{Vincent Starck\thanks{I gratefully acknowledge financial support from the European Research Council (Starting Grant No. 852332). I thank Susanne Schennach for many helpful comments; I am also grateful to participants at various conferences for fruitful discussions that help improve this paper.
}~\thanks{%
V.Starck@lmu.de}\\ LMU Munich}
\date{\today}

\newcommand{\meangG}{\frac{1}{N} \sum_{g=0}^G}
\newcommand{\sumin}{\sum_{i=1}^n}

\newcommand\equaldef{\mathrel{\overset{\makebox[0pt]{\mbox{\normalfont\tiny\sffamily def}}}{=}}}
 
\DeclareMathOperator*{\argmin}{arg\,min} 

\newcommand{\indep}{\perp \!\!\! \perp}
\DeclarePairedDelimiter\floor{\lfloor}{\rfloor}

\begin{document}

\maketitle\thispagestyle{fancy}

\begin{abstract}

I study peer effects that arise from irreversible decisions in the absence of a standard social equilibrium. I model a latent sequence of decisions in continuous time and obtain a closed-form expression for the likelihood, which allows to estimate proposed causal estimands. The method avoids linear-in-means regression by modeling the (possibly unobserved) realized direction of causality, whose probability is identified. I provide identification and estimation results under two settings, several networks and one large network, while allowing for various forms of peer effect heterogeneity. Under (strong) data requirements, it is possible to separate endogenous, contextual, and correlated effects while allowing for full heterogeneity and maximum likelihood methods where parameters lend themselves to standard inference.


\end{abstract}

\section{Introduction}

Analyzing peer effects is notoriously difficult. In a seminal paper, \citet{manski1993identification} discusses the reflection problem that arises when one runs a regression on the conditional expectation, which induces restrictions on the regression coefficients that lead to tautological models or identification issues. \medskip

In applications that feature small group or friendship, the channel of influence plausibly operates through actual outcomes, which leads to Spatial Autoregressive (SAR) or linear-in-means (LIM) models in the outcomes. They alleviate reflection problems insofar as they allow for identification and estimation of peer effects \citep{bramoulle2009identification, de2010identification, blume2015linear, martellosio2022non}, \textit{e.g.}, through the use of non-overlapping peer groups that instruments away the simultaneity bias. \medskip 

However, inference about the effect of peers using SAR models can be challenging. \citet{manski1993identification} points out difficulties in identifying and estimating peer effects that transcend reflection regression issues. Identification can be tenuous in SAR models as it is based on functional form restrictions that are particularly sensitive to\textit{, e.g.}, measurement error or mechanical relationships between individual outcomes and group means \citep{gibbons2012mostly, angrist2014perils}. Although identification is typically the rule when the network is known \citep{blume2015linear}, identification is dependent on the mechanics of the model through its reduced form. Moreover, reliable inference about peer effects comes with caveats even in the absence of identification problems \citep{hayes2024peer, wang2025weak}. \medskip

Frameworks for analyzing peer effects beyond linear models are scarce but often desirable. As summarized in \citet{sacerdote2014experimental}, ``researchers have shown that linear-in-means model of peer effects is often not a good description of the world, although we do not yet have an agreed-upon model to replace it.'' \citet{boucher2024toward} provide a recent breakthrough that extends the response class to contain mean, maximum, and minimum outcome in situations of equilibrium. \medskip

Focusing on the case of irreversible decisions, I develop a new framework by modeling the latent sequence of decisions in continuous time. In this case, the standard notion of social equilibrium\footnote{The solution for the conditional expectation or the vector of outcomes implied by the linear-in-means specification.} is inadequate as people opt in over time according to their characteristics and shocks, then cannot subsequently adjust their outcomes. This setup offers a natural framework to discuss causality, as it breaks the simultaneity by separating first-movers that potentially generate a causal reaction from the subjects of influence. I formalize the counterfactual framework using potential outcomes, and define causal peer effect parameters that lend themselves to straightforward inference via maximum likelihood. \medskip 

In this context, an important object is the order in which individuals select into the absorbing state. When unknown, it is potentially an object of interest. Although the realized order cannot be identified, the probability of any order can be identified as a function of individuals' network and characteristics; I provide a framework to estimate those probabilities. Conversely, when the order is (partially) known, it carries identifying power that is useful for identifying and estimating peer effects without restricting heterogeneity, functional form, or the presence of contextual and correlated effects. \medskip

As we distinguish between sources and recipients of peer influence, exploring peer-effect heterogeneity can be more fruitful, as it provides parameters that are more natural to interpret. Heterogeneity in peer effects is an important, but under-explored topic (recent work on the subject includes \citet{mogstad2024peer}). An important source of heterogeneity may be that the influence of $i$ on $j$ need not be the same as that of $j$ on $i$, whence the relevance of orderings. For instance, people may react differently upon observing a popular or highly educated peer opting in than this peer would in the converse situation. It is also likely that some characteristics make people move first with high probability. \medskip 

I analyze identification under two main regimes: one large network and many networks. Large networks with observed ordering are found to be fully identified, \textit{i.e.}, peer effects of heterogeneous forms are identified separately from influence of covariates, contextual effects, and correlated effects under (strong) data requirements. Identification stems from information about the order of moves and is not tied to functional form restrictions; no aspect of heterogeneity needs to be restricted. \medskip 

In the many, small network case, identification is more tenuous because correlated effects may create incidental parameters, which is especially problematic if those effects can interact with any other covariate or peer effect strength. In this case, they must be handled by a combination of functional form restrictions, distributional assumptions, and extra information. In the absence of correlated effects, identifying heterogeneous, nonlinear peer effects and the influence of covariates is feasible. \medskip

The paper contributes to the literature on peer effects, in particular identification and estimation issues \citep{manski1993identification, bramoulle2009identification,angrist2014perils,blume2015linear,hayes2024peer}. In addition, the results on orderings make connections to the literature on targeting and diffusion \citep{banerjee2013diffusion, he2018measuring} while the method may also be useful to analyze staggered treatment adoption \citep{shaikh2021randomization, athey2022design} by relaxing the common assumption that treatment adoptions arise independently.

\section{Causality, stochastic process, and likelihood}

\subsection{Setup and causal peer effect parameters}

I focus on irreversible\footnote{This can be because the action cannot be undone (\textit{e.g.}, vaccination), is too costly to reverse, or because the focus is on first-time events.} decisions: there is an initial default state, labeled $0$, and the decision to opt-in leads to an absorbing state, labeled $1$. For instance, the outcome $y$ might represent vaccination status, technology adoption, retirement, migration decision, etc. The goal is to model and estimate peer effects, \textit{i.e.} how decisions of peers alter an individual's probability of opting in. The peers are described by a network; their importance is given by a weighting matrix $W$. Although it is naturally interpreted as a row-normalized matrix giving positive weights to each individual's peers, the results apply to general matrices provided that $W$ is bounded in row and column sums. \medskip

The adoption time of an individual $i$, denoted by $T_i$, is a random variable that depends on individuals' characteristics and their expectations. Observing a peer opting in modifies the likelihood that an individual does so as well. This can be due to conformity, information transmission, or other types of social influence. This implies that there is a change in the distribution of the adoption time of individual $i$, $T_i$, upon observing the adoption of a peer. \medskip

Adapting the potential outcome notation \citep{neyman1923application, rubin1974estimating} to the current setup\footnote{Potential outcomes have been used to formalize causality in peer effects in different contexts. The literature on interference (``exogeneous peer effects'' or sometimes spillovers), in which one's outcome depends on neighbors' treatment, addresses (versions of) the issue and identifies direct and indirect effects under various relaxations of SUTVA \citep{toulis2013estimation, sofrygin2016semi, aronow2017estimating, arpino2017implementing, liu2019doubly, forastiere2020identification, jackson2020adjusting, sanchez2021spillovers, huber2021framework}. Potential outcome that depends on peer's outcome (\textit{e.g.}, in \citet{egami2024identification}) have been the object of less discussion. To my knowledge, the discussion of peer effects through potential outcomes in the time dimension is new.}, the adoption time is represented by $T_i(\tau)$ for $\tau \in \mathds{R}^{n-1}$: the adoption time of individual $i$ depends on the adoption times of other individuals. This dependence carries over to the outcomes $\mathds{1}_{T_i \leq t}$, which are observed for $t=S>0$: $y_i \equaldef \mathds{1}_{T_i \leq S}$. \medskip 

In general, the resulting change in the distribution of outcomes following a change in $\tau$ can be interpreted as a peer effect. The causal impact can be captured by parameters that summarize the effect of a change in $\tau$ on the distribution. \medskip

A parameter of interest is the analog of the average treatment effect: the mean change in the outcome following a change in $\tau$:

\begin{equation}
    \delta(\tilde{\tau}, \tau) \equaldef \mathds{E}[Y_i(\tilde{\tau})-Y_i(\tau)] = \mathds{P}[Y_i(\tilde{\tau})=1]-\mathds{P}[Y_i(\tau)=1]
\end{equation}

For instance, one could be interested in $\tilde{\tau}=\vec{0} \in \mathds{R}^{n-1}$ and $\tau=\infty e_i$\footnote{$e_i$ is a vector whose only nonzero entry is a $1$ in place $i$. I adopt the convention $0 \cdot \infty = 0$.} or $\tau= \vec{\infty}$ so that the parameter describes the change in adoption probability induced by the initial adoption of one or all peers compared to them never adopting. \medskip

We could also be interested in counterfactual effects such as the expected adoption time, $\mathds{E}[T_i]$, and various conditional versions, or the expected time before a fraction of the population opts in. 

\begin{remark}[Beyond binary outcomes] The process can be extended to account for more complex decisions. For instance, there could be a set of absorbing states to select into or there could be a gradation in behavior (say, $y_i \in \mathds{N}$ is non-decreasing over time). The analysis can be extended to such frameworks at the expense of additional notation and suitable regularity conditions. \medskip
\end{remark}

\subsection{Stochastic process and unconfoundedness}\label{stochastic_process}

Individual adoption is modeled with the following continuous-time stochastic process: 
\begin{definition}[Stochastic process]
    Let $T_i^1$ be a set of random variables over $\mathds{R}^+$. When a first individual, say $j$, opts in (\textit{i.e.,} $j = \arg \min_i (T_i^1)$), she withdraws and the remaining random variables are updated to $T_i^{2}$, whose distribution may depend on the time elapsed and the event that $j$ opted in.  
    The process then goes on with the updated distributions, and so on. The time to adoption is therefore $T_i = \sum_{k=1} T_i^k$, where the sum ranges from $1$ to the round where $i$ comes first. The outcomes are observed at time $S$: $y_i = \mathds{1}_{T_i \leq S}$. 
\end{definition} 


The process decomposes the arrival time $T_i$ of each individual $i$ into a collection $T_i^* = (T_i^1, \ldots, T_i^n)$ of \textit{latent partial times}. 
This process is quite general in terms of the dynamics it allows. It basically only assumes the arrow of time, ruling out feedback from the future. \medskip

Peer effects occur when the updated distributions do not coincide with the previous distribution. Latent partial times thus incorporate distributional changes due to peer effects, but are usually correlated because characteristics influence both the choice of peers and the distribution of times, inducing, \textit{e.g.}, homophily bias \citep{shalizi2011homophily}. Controlling for these characteristics can restore independence:

\begin{assumption}[Independence of latent partial times]
    The latent partial times are conditionally independent: $\forall i, T_{-i}^{k+1} \indep T_{i}^{k+1} \vert X, W, \mathcal{F}_k$
\end{assumption}
where $\mathcal{F}_k$ is the relevant filtration (collecting the previous $\argmin_i T_i^k$ and $\min_i T_i^k$, \textit{i.e.}, the identity of previous movers and their (partial) times. It also satisfies $\mathcal{F}_0 = \emptyset$).


The independence of latent partial times is related to a notion of unconfoundedness. Note that the potential outcomes can be expressed in terms of the latent partial times $T_i^{k}, k=1, \ldots, $ \textit{via} 
\begin{equation}
    T_i(\tau) = \sum_{k} \left(\mathds{1}(T_i^{k} < \tau_{(k)} - \tau_{(k-1)}) T_i^{k} + \mathds{1}(T_i^k \geq \tau_{(k)} - \tau_{(k-1)}) \ (\tau_{(k)} - \tau_{(k-1)})\right)
\end{equation}
\noindent where $\tau_{(0)} \equaldef 0$, the sum ranges from $1$ to $\argmin_k \{T_i^k < \tau_{(k)} - \tau_{(k-1)}\}$, and the latent partial times are based on $\tau$. \medskip

Then, unconfoundedness can be stated as follows. 

\begin{assumption}[Unconfoundedness of timings]
    Potential outcomes are independent of times of arrival: $\forall i, k, \tau, \ T_{-i}^{k+1} \indep T_i(\tau) \vert X, W, \mathcal{F}_k$
\end{assumption}
\noindent where $T_{-i}$ is the set of times of all individuals but $i$. \medskip

\noindent Then, the following result follows immediately from the formulation of potential outcomes:

\begin{proposition}
    Unconfoundedness of timings is equivalent to independence of latent partial times.
\end{proposition}

Unconfoundedness, or equivalently independence of latent partial times, ensures that some (changes in) parameters have a causal interpretation. It states that network links are not predictive of unobserved factor which influences the time when people opt in. This requires controlling for variables that affect both network formation and the probability of participation, which is a strong assumption, but similar to frequently invoked exogeneity assumptions, \textit{e.g.}, \citet{bramoulle2009identification}. \medskip

Although beyond the scope of this paper, modeling network formation to identify or control for latent variables \citep{goldsmith2013social, graham2017econometric, auerbach2022identification, starck2025improving} can help in the event of unobserved confounders. When feasible, randomization of peer groups provides a natural way to ensure unconfoundedness. \medskip

\subsection{Likelihood}

Absent information about the times when people opt in, the latent distributions are not identified. To make the model tractable and easy to interpret, I specify the distribution to be exponential and assume that the rates depend on the peers who previously opted in:
\begin{assumption}
    $T_i^{k+1} \vert X, W, \mathcal{F}_{k} \sim \mbox{Exp}(\lambda_i(X, W, \mathcal{F}_{k}))$
\end{assumption}

I focus on Exponential distributions for two reasons. First, exponential waiting times arise automatically under the assumption of a constant probability per unit of time, a natural point of departure. Second, exponential distributions are particularly attractive from an analytical standpoint and ensure tractability. Due to the memorylessness property, conditioning on elapsed time is irrelevant. If we assume that the peer effects depend on the identity of the previous movers (regardless of their order), the relevant filtration consists only of a set with the identity of the previous movers. In the absence of peer effects, the rates simply do not evolve: $\lambda_i^{+\mathcal{F}_k} = \lambda_i$ for all $i, k$.
In what follows, I let $\lambda_i^{+\mathcal{F}_k} \equiv \lambda_i(X, W, \mathcal{F}_{k})$ with $\lambda_i \equiv \lambda_i^{+\emptyset}$. \medskip

Importantly, exponential rates can be left unrestricted as functions of covariates. Given a distribution of the outcomes, they are nonparametrically identified under few conditions. As such, the exponential specification provides a convenient framework in which peer effects are easy to interpret, while allowing for considerable heterogeneity. Under the information structure considered here (outcomes at time $S$, possibly order of adoptions), the exponential specification provides a fit to the latent time dynamics that facilitates interpretation without letting peer effect identification depend on functional form. 
The assumption has more bite when time dynamics is explicitly used for identification or when time analysis is of interest (\textit{e.g.}, diffusion analysis, generation of counterfactuals, etc.), because the exponential distribution over time then has an influence on estimates. \medskip

The exponential rates capture the heterogeneity across individuals and their changes over time reflect the influence of peers. Their changes can be directly interpreted as peer effects insofar as they capture the change in the shape of the distribution and the measure average reduction in adoption times. They also translate into practical formulae to compute parameters of interest.

\begin{example}
    Consider the change in probability of adoption:
\begin{equation}
    \mathds{E}[Y_i(\vec{0}) - Y_i(\vec{\infty})\vert X] = e^{-\lambda_i S} (1-e^{-(\lambda_i^{+} - \lambda_i) S)})
\end{equation}
where $+$ corresponds to the final level of peer effects -- all peers have previously opted in. \medskip

\noindent To the first-order, this simplifies to
\begin{equation}
    e^{-\lambda_i S} (\lambda_i^{+} - \lambda_i) S
\end{equation}

The first term, $e^{-\lambda_i S}$, is the baseline probability of not opting in before time $S$, while $(\lambda_i^{+} - \lambda_i)$ reflects the change in exponential rates induced by the prior adoption of friends. 
\end{example}

The following examples consider simple homogeneous cases to illustrate the framework and build intuition for identification conditions, which are formalized later.

\begin{example}\label{two_people}

Consider two (connected) individuals ($i=1, 2)$ and homogeneous rates $\lambda, \lambda^+$. The probabilities for the four possible outcomes are derived in the appendix. They are given by
\begin{align*}
    \begin{split}
        & p_{00} \equaldef \mathds{P}[y_1=y_2=0\vert W_{12} =1] = e^{-2 \lambda S} \\
        & p_{10} \equaldef \mathds{P}[y_1=1, y_2=0\vert W_{12} =1] = \lambda e^{-\lambda^+ S} g(2 \lambda - \lambda^+)\\
        & p_{01} \equaldef \mathds{P}[y_1=0, y_2=1\vert W_{12} =1] = \lambda e^{-\lambda^+ S} g(2 \lambda -\lambda^+) \\
        & p_{11} \equaldef \mathds{P}[y_1=y_2=1\vert W_{12} =1] = 1 - p_{00} - p_{10} - p_{01}
    \end{split}
\end{align*}
where $g(\lambda) = \frac{1-e^{-\lambda S}}{\lambda}$ if $\lambda \neq 0$ and $g(0) = S$. \medskip

The probabilities are identified is one observes a sequence of independent draws of such pairs. In this case, the rates are identified since they can be recovered from the probabilities: $\lambda = -\frac{\ln(p_{00})}{2S}$ while $\lambda^+$ solves $\frac{\lambda S(p_{00}-e^{-\lambda^+ S})}{\ln(p_{00}) + \lambda^+ S} =p_{10}$, where the left-hand side is strictly decreasing. \medskip



Although it is not possible to identify the identity of the first mover -- the individual who may have exerted a peer effect on the other -- when $y_1=y_2=1$, it is possible to (i) estimate the peer effect strength and (ii) determine the probability of an individual moving first, which provides a probability for each direction of causality.
\end{example}

\begin{example}\label{complete_network_example}
Consider a complete network of size $n$, \textit{i.e.}, the adjacency matrix corresponds to $\iota \iota' - I$ where $\iota$ is a vector of ones. People have baseline rates $\lambda$ that are updated to $\lambda^{+^k}$ when $k$ people have opted in. \medskip

The outcomes $y_i \in \{0, 1\}$ only inform the fraction of people who opted in, which is insufficient to identify $\{\lambda^{+^k}, k=0, \ldots, n-1\}$ as only the $k$-th mover provides a signal about $\lambda^{+^k}$. In this regime, positive peer effects cannot be distinguished from stronger baseline rates. Because of the symmetry and perfect connectivity, knowledge about the order of moves is also uninformative. \medskip

The complete network case thus requires more information, but can still be identified under additional structure. For instance, if $\lambda = e^{\beta}$ for some $\beta \in \mathds{R}$ and $\lambda^{+^k} = e^{\beta+\frac{k}{n-1} \delta}$ and the times of adoption of early adopters are known, then $\lambda$ is identified from the behavior of the density around the origin.\footnote{In this case, identification relies on the exponential specification because the identifying power of the time of adoptions needs to be used.} Given the knowledge of $\lambda$, $\delta$ is identified from the share of adopters. \medskip

The source of non-identification is the lack of repeated signals about rates of a given order when everyone is connected to everyone. This suggests that one can rely on sparser networks to secure identifying information that does not rely on time data or further rate restrictions. This intuition is developed in the next Section. In particular, when the \textit{order} of adoption is known, then the rates can be shown to be identified when degrees are bounded. \medskip



\end{example}

The likelihood induced by the stochastic process is available in closed form. This result is the object of the next theorem. 

\begin{theorem}\label{likelihood_theorem}
If the latent partial times are independent, then the log-likelihood for the exponential specification takes the form
    \begin{equation}
        l_N \equaldef N^{-1} \ln\left(\sum_{p \in \mathcal{P}} G! \vert \mathcal{P} \vert^{-1} \left(\prod_{i=1}^G \lambda_{p_i}^{+\{p_1, \cdots, p_{i-1}\}}\right) \sum_{g=0}^{G} \frac{e^{-{\Lambda}_{p_g} S}}{\prod_{h \neq g} {\Lambda}_{p_h} - {\Lambda}_{p_g}} \right)
    \end{equation}
    where $\mathcal{P}$ contains permutations of the $G$ people who opted in, to which the remaining people are appended in any order, and $\Lambda_g \equaldef \sum_{i=g+1}^N \lambda_{p_i}^{+\{p_1, \cdots, p_g\}}$.
\end{theorem}

The following example shows that the process generalizes i.i.d. exponential draws. The heterogeneity in the rates induces different distributions between individuals, while the peer effects reflected in $\lambda^+ \neq \lambda$ create spatial dependence. 

\begin{example}
Suppose that there is no heterogeneity nor peer effects: $\lambda_i^{+\mathcal{F}} \equiv \lambda$ for all $i$ and any $\mathcal{F}$. Then $\Lambda_g = \lambda (N-g)$ and the likelihood any ordering becomes 

\begin{align*}
    \begin{split}
    G! \prod_{i=1}^G \lambda_i^{+\{(1), \cdots, ({i-1})\}} \sum_{g=0}^{G} \frac{e^{-\Lambda_g S}}{\prod_{h \neq g} \Lambda_h - \Lambda_g} & = G!  \lambda^G \sum_{g=0}^{G} \frac{e^{- \lambda (N-g)S}}{\prod_{h \neq g} \lambda (N-h)-\lambda (N-g)} \\
    & = G!  e^{-\lambda (N-G) S} \sum_{g=0}^{G} \frac{e^{- \lambda (G-g)S}}{\prod_{h\neq g} g-h} \\
    & = e^{-\lambda (N-G) S} \sum_{g=0}^{G} G! \frac{e^{- \lambda (G-g)S}}{g! (G-g)!} (-1)^{G-g} \\
    & = e^{-\lambda (N-G) S} (1-e^{-\lambda S})^G,
    \end{split}
\end{align*}
which is the likelihood for $G$ (out of $N$) i.i.d. exponentially distributed variables falling below a cutoff $S$. \medskip

Therefore, probabilities reduce to a standard exponential race with independent draws when the rates do not vary. Changes in exponential rates are a measure of dependence and social interaction; they reflect peer effects.  
\end{example}

\begin{remark}[2 people example]
    The probabilities in Example \ref{two_people} can be directly obtained from the theorem. For instance, noting $\Lambda_{p_1} = \lambda_1+\lambda_2$ and $\Lambda_{p_2} = \lambda_{p_2}^+$, one can verify 
    \begin{align}
        \begin{split}
            \mathds{P}[y_1=y_2=1\vert W_{12} =1] = 1 & + \frac{\lambda_1 \lambda_2^+ e^{-(\lambda_1+\lambda_2)S}}{(\lambda_1+\lambda_2-\lambda_2^+)(\lambda_1+\lambda_2)} - \frac{\lambda_1 e^{-\lambda_2^+ S}}{\lambda_1+\lambda_2-\lambda_2^+} \\
            & + \frac{\lambda_2 \lambda_1^+ e^{-(\lambda_1+\lambda_2)S}}{(\lambda_1+\lambda_2-\lambda_1^+)(\lambda_1+\lambda_2)} - \frac{\lambda_2 e^{-\lambda_1^+ S}}{\lambda_1+\lambda_2-\lambda_1^+} \\
            & = 1 - p_{00} - p_{10} - p_{01}
        \end{split}
    \end{align}
    
\end{remark}

\begin{remark}[Computation]
Although summing over all permutations can lead to an impractical computational burden, the computational cost can be lowered to practical levels. 
First, the likelihood factorizes based on the components of $W$, whose size can be much smaller (classrooms, villages, connected components of friends, etc.). Second, permutations are based on the number of adopters, rather than the size of each component of the network. Hence, the complexity of permutations is substantially reduced, especially when partial information about the ordering is available. \medskip

Finally, approximations can further reduce the number of permutations. In particular, the average over permutations can be estimated by a random sample of permutations by the law of large numbers. The likelihood, score, and Hessian can all be approximated using this technique. \medskip


\end{remark}

\subsection{Identification}

The sample comprises outcomes ($y_i, i=1, \ldots, n$), covariates $(x_i, i=1, \ldots, n)$, and a weighting matrix $W$ ($W_{ij} > 0$ if $i$ and $j$ are peers) that determines the peer group of each individual. \medskip

Identification relies on our ability to separate information about cross-sectional variation in rates and social influence effects. In the complete network of Example \ref{complete_network_example}, this is not possible without further information because once a single person opts in, everybody else is subject to social influence and there is no additional information about baseline rates. Most real-life networks, however, are much sparser or exhibit a block structure due to a sampling scheme that targets villages, classrooms, etc. I show how this provides the necessary information for identification under the two regimes: the network consists of components or blocks which do not interact, or are sufficiently sparse. In the latter case, I make use of the assumption that degrees are bounded, which has been invoked in the literature to consider sparser networks (\textit{e.g.}, \citet{de2018identifying}, though it can be relaxed to a slow degree growth rate.

\begin{lemma}\label{identification_lemma}
    Let $x_i$ be composed of discrete and continuous variables, and let the rates be continuous in the continuous variables. 
    Assume that latent partial times are independent and that the exponential specification holds with $\lambda_i^{+\mathcal{F}} \in A \subseteq \ ]0, \infty[$, for all $i$ and set $\mathcal{F}$. Then, \\
    (i) If the network has a block structure with blocks $b=1, \cdots, B$, of size $N_b \leq \overline{N}$ for some $\overline{N} \in \mathds{N}$, then $\lambda_i^{+\mathcal{F}}$ is identified whenever $\mathds{P}[W_{b;ik}>0, \forall k \in \mathcal{F}\vert X] > 0$, $\mathds{P}[Y_b=y_b\vert W_{b;ik}>0, \forall k \in \mathcal{F} \vert X] > 0$, and the number of blocks grows large. \\
    (ii) If the order of adoption is known, the degrees are bounded, \textit{i.e.}, $\sup_{i} d_i \leq \overline{d} < \infty$ where $d_i = \sum_{j=1}^N \mathds{1}(W(i, j) > 0)$, and $A$ is compact, then $\lambda_i^{+\mathcal{F}}$ is identified for all $i$ and any set of people $\mathcal{F}$ whose characteristics make a connection with $i$ possible with positive probability. 
\end{lemma}

\begin{remark}[Identification from panel information]
Identification using the order of adoption can be extended to setups where the order is only partially known. For instance, monitoring the activations at regular times provides a partial order that secures identification as the frequency grows.
\end{remark}

\begin{remark}[Identification of order probabilities]
Identification of the family of rates implies that the probability of any sequence of adoption is also identified. Hence, although it is not possible to recover the identity of the first movers or the realized order of adoptions, it is possible to assign a probability to such events. 
\end{remark}

\begin{remark}[Contextual and correlated effects]

Contextual peer effects can be straightforwardly allowed for in the large network case. Within a large network, the absence of correlated effects hinges only on the (strong) assumption of no unobserved confounders. Thus, under the maintained unconfoundedness condition, information about orderings is sufficient to identify peer effects in full heterogeneity while separating them from contextual and correlated effects. \medskip

In general, note that no restriction has been placed on the way the specific network heterogeneity may interact with covariates or peer effects. It is thus unsurprising that identification generically fails: separating endogenous effects from contextual or correlated effects requires further information. In the large network case, this can take the form of the order of adoption. With several networks, which induce incidental parameters, this would require functional form restrictions (in the spirit of the SAR reduced form, for instance), distributional assumptions, or partial information about the order of adoptions combined with other restrictions. 
\end{remark}

\section{Asymptotic Theory}

Although rates are nonparametrically identified, it is crucial to introduce more parsimonious specifications. This avoids severe curse-of-dimensionality issues due to the large number of rates and the dependence on possibly many continuous covariates and simplifies interpretation. \medskip

A natural model specifies the baseline rates as $\lambda_i = g(x_i, \beta)$ for a positive link function determined by a finite-dimensional vector of parameters $\beta$, and lets the updated rates be obtained by a scaling factor $\delta$. For instance, a simple specification is
\begin{equation}\label{parametric_spec_for_lambdas}
    \ln(\lambda_i^{+_{\mathcal{F}}}) = x_i^' \beta + \sum_{j \in \mathcal{F}} W(i, j) \delta.
\end{equation}

Such a specification reduces the dimensionality of the problem to $(\mbox{dim}(x_i)+1)$, ensures positive rates, and lets peer effects be described by a parameter $\delta$ that reflects how rates are scaled when the members of the peer group opt in. $\delta$ controls the strength of spatial dependence. \medskip

    To the first-order in $\delta$, we have
\begin{equation}
    \mathds{E}[Y_i(\vec{0})-Y_i(\vec{\infty})\vert X] \approx \delta S \lambda_i e^{-\lambda_i S}
\end{equation}
so that $\delta$ acts a scaling factor in the change in the probability of adoption induced by the peer group adopting at the onset. \medskip

It is easy to relax the parametric specification in various directions. The relevant direction is an applied choice that is likely to differ between applications.
For instance, suppose that a researcher is analyzing peer effect in vaccine uptake. They might conjecture that people respond differently depending on the level of education of their peers and that the marginal peer effect is decreasing because the first few movers provide most of the information transmission or reassurance about safety. A natural way to explore heterogeneity in peer effects is then to add interaction terms such as $\sum_{j \in \mathcal{F}} W(i, j) x_j$, nonlinear terms in $\sum_{j \in \mathcal{F}} W(i, j)$ such as powers, or $p$-norms in the spirit of \citet{boucher2024toward}. \medskip

For a given parameterization of the rates in terms of a parameter $\theta$ (for instance, $\theta = (\beta', \delta)'$ in (\ref{parametric_spec_for_lambdas})), estimation proceeds by maximum likelihood. Asymptotics can be obtained in two frameworks: multiple networks and one large network. \medskip 

In the first case, I consider a sequence of networks with blocks or components of size $N_b$, $b=1, \ldots, B$. With independent blocks, the likelihood factorizes as $l \equaldef \ln(\mathds{P}[Y=y]) = \sum_{b=1}^B l_b$, where $l_b$ is the log-likelihood of block $b$. The logarithmic likelihood of each block is given by the formula established in the theorem of the previous section. The score and Hessian are easily computed in closed form. The details are provided in the Appendix. In the second case, the network consists of a single component whose size grows and a key ingredient comes from knowledge of $\mathcal{F}$, as in the identification condition. In this case, I proceed conditionally on the covariates and the network structure. \medskip

In both cases, the estimator $\hat{\theta}$ inherits the usual properties of maximum likelihood estimators: it is consistent and asymptotically normal under regularity conditions. This implies consistency and asymptotic normality of the family of rates. Formally, 

\begin{theorem}[Consistency]\label{consistency_theorem}
Let the data-generating process be described by the stochastic process of Section \ref{stochastic_process} and that the family of rates is identified and belongs to the interior of a compact set.  
Then, (i) if blocks are drawn independently with $W_b \sim \mathcal{D}_W$ such that $N_b<\overline{N}$, the maximum likelihood estimator is consistent with
$\hat{\lambda}_i^{+\mathcal{F}} \overset{p}{\longrightarrow} \lambda_i^{+\mathcal{F}}$, for any $\mathcal{F} \subseteq \{1, \ldots, i-1, i+1, \ldots, N_b\}$, as $B \rightarrow \infty$.
Alternatively, (ii) if the size of the network grows large and condition (ii) of Lemma \ref{identification_lemma} is satisfied, the maximum likelihood estimator is consistent with
$\hat{\lambda}_i^{+\mathcal{F}} \overset{p}{\longrightarrow} \lambda_i^{+\mathcal{F}}$, for any $\mathcal{F}$, as $N \rightarrow \infty$.
\end{theorem}

Moreover, 

\begin{theorem}[Asymptotic Normality]\label{normality_theorem}
Assume that part (i) of the assumptions of the Consistency theorem hold and that the derivatives $\partial_\theta \lambda_i^{\mathcal{F}}$ are bounded. Then, the maximum likelihood estimator is asymptotically normal as $B \rightarrow \infty$ with 

\begin{equation}
    \sqrt{B} (\hat{\theta}-\theta) \overset{d}{\longrightarrow} \mathcal{N}\left(0, - H^{-1}\right)
\end{equation}

Alternatively, assume that part (ii) of the assumptions of the Consistency theorem hold and that the derivatives $\partial_\theta \lambda_i^{\mathcal{F}}$ are bounded. Then, the maximum likelihood estimator is asymptotically normal as $N \rightarrow \infty$ with 
\begin{equation}
    \sqrt{N} (\hat{\theta}-\theta) \vert X, W \overset{d}{\longrightarrow} \mathcal{N}\left(0, - H^{-1}\right).
\end{equation}
Here, the Hessian is given by 
\begin{equation}
    H \equaldef - \int_0^c \frac{\partial_{\theta \theta'} r(z)}{r(z)+z^*} \mbox{d}z + \int_0^c \frac{\partial_{\theta} r(z) \partial_{\theta'} r(z)}{(r(z)+z^*)^2}  \mbox{d}z - \frac{\int_0^c \frac{\partial_{\theta} r(z)}{(r(z)+z^*)^2} \mbox{d}z \ \int_0^c \frac{\partial_{\theta'} r(z)}{(r(z)+z^*)^2} \mbox{d}z}{\int_0^c \frac{1}{(r(z)+z^*)^2} \mbox{d}z},
\end{equation}
where $r$ is a limiting function for a transformation of the rates $\Lambda_g$, $c = \lim_{N\longrightarrow \infty} \mathds{E}[G/N]$, $z^*$ satisfies a stationarity condition, as described in the Appendix. 

Finally, by the Delta method, the rates obey
\begin{equation}
    \sqrt{N} (\hat{\lambda}_i^{+\mathcal{F}} - \lambda_i^{+\mathcal{F}}) \vert X, W \overset{d}{\longrightarrow} \mathcal{N}\left(0, - \partial_{\theta'} \lambda_i^{+\mathcal{F}} H^{-1} \partial_{\theta} \lambda_i^{+\mathcal{F}}\right)
\end{equation}
\end{theorem}

These theorems allow for standard inference about the underlying parameters and the rates. \medskip

\bibliographystyle{aea}
\bibliography{References}

\section*{Appendix A: Proofs and Further results}

\subsection*{Probabilities in Example \ref{two_people}}

\begin{proof}
The result is shown for possibly different rates $\lambda_1, \lambda_2$. 
Consider the probability $\mathds{P}[y_1 =1, y_2 = 0]$. This corresponds to
\begin{align}
    \begin{split}
        \mathds{P}[T_1 \leq S, T_2 > S] & = \int_0^S p[T_1=t, T_2 > S] \ dt \\
        & = \int_0^S \lambda_1 e^{-\lambda_1 t} \ p[T_2> S\vert T_1 = t] \ dt \\
        & = \int_0^S \lambda_1 e^{-\lambda_1 t} \ p[T_2^1> t\vert T_1 = t] \ p[T_2^2> S-t\vert T_1 = t] \ dt \\
        & = \int_0^S \lambda_1 e^{-\lambda_1 t} \ p[T_2^1> t] \ p[T_2^2> S-t\vert T_1 = t] \ dt \\
        & = \int_0^S \lambda_1 e^{-\lambda_1 t} e^{-\lambda_2 t} e^{-\lambda_2^{+} (S-t)} \ dt
    \end{split}
\end{align}
where densities are denoted by $p$. 

If $\lambda_1 + \lambda_2 - \lambda_2^+ = 0$, this is simply $\lambda_1 S e^{-\lambda_2^+ S}$. If $\lambda_1 + \lambda_2 - \lambda_2^+ = 0$, then integrating yields $\lambda_1 e^{-\lambda_2^+ S} (1-e^{-(\lambda_1+\lambda_2-\lambda_2^+)S})/(\lambda_1+\lambda_2-\lambda_2^+)$. 

Then, $\mathds{P}[y_1 =0, y_2 = 1]$ follows by symmetry, $\mathds{P}[y_1 =0, y_2 = 0] = e^{-(\lambda_1+\lambda_2)S}$ is trivial, and $\mathds{P}[y_1 =1, y_2 = 1]$ is deduced by the fact that the four probabilities sum up to $1$.

\end{proof}

\subsection*{Proof of Theorem \ref{likelihood_theorem}}
\begin{proof}

The log-likelihood is given by 
\begin{equation}
    \ln(\mathds{P}[Y_1 = y_1, \ldots, Y_N =y_N \vert X, W])
\end{equation}

\noindent Assume without loss that the individuals who opted in correspond to the first $G$ observations. Then, summing over different orderings,
\begin{align*}
    & \mathds{P}[y_1 = \ldots = y_G = 1 ; y_{G+1} = \ldots = y_N =0] \\
    & = \sum_{p \in \mathcal{P}} \mathds{P}[y_1 = \ldots = y_G = 1 ; y_{G+1} = \ldots = y_N =0; y_{G+1}, \cdots, y_N > S; T_{p_1} < \ldots < T_{p_G}]
\end{align*}
where the sum is over all permutations (with generic element $p = (p_1, \ldots, p_G)$) of the $G$ first arrivals. \medskip

Consider the representative term in the sum when individual $i$ is the $i$-th to opt in. Using the independent of partial times and the exponential specification and following steps similar to the 2-individual examples, this term is given by
\begin{align*}
    \begin{split}
    & \int_0^S  \int_{t_1^1}^\infty \cdots \int_{t_1^1}^\infty \prod_{i=1}^N \lambda_i e^{-\lambda_i t_i^1} \quad \int_0^{S-t_1^1} \cdots \int_{t_2^2}^\infty \prod_{i=2}^N \lambda_1^{+_1} e^{-\lambda_i^{+_1} t_i^2} \cdots \\
    & \int_0^{S-\sum_{g=1}^{G-1} t_g^g} \cdots \int_{t_G^G}^\infty \prod_{i=G}^N \lambda_i^{+_1 \cdots +_{G-1}} e^{-\lambda_i^{+_1 \cdots +_{G-1}} t_i^G} \\ 
    & \int_{S-\sum_{g=1}^G t_g^g}^\infty \cdots \int_{S-\sum_{g=1}^G t_g^g}^\infty \prod_{i=G+1}^N \lambda_i^{+_1 \cdots +_G} e^{-\lambda_i^{+_1 \cdots +_G} t_i^{G+1}} \\
    & dt_1^1 \cdots dt_N^1 dt_2^2 \cdots dt_N^2 \cdots dt_G^G \cdots dt_N^G \cdots dt_1^{G+1} \cdots dt_N^{G+1}
    \end{split}
\end{align*}
which, after some algebra, reduces to 
\begin{equation}
    e^{- \sum_{i=G+1}^N \lambda_i^{+_1, \cdots, +_G} S} \left(\prod_{i=1}^G \lambda_i^{+_1, \cdots, +_{i-1}} \right) I_{\{c_i, i=1, \cdots, G\}}
\end{equation}
with $I_{\{h_i, i=1, \cdots, H\}} \equaldef \frac{1}{\prod_{g=1}^G \dot{\Lambda}_g} + (-1)^G \sum_{g=1}^G \frac{1}{\prod_{h\neq g} (\dot{\Lambda}_g - \dot{\Lambda}_h)} \frac{e^{-\dot{\Lambda}_g S}}{\dot{\Lambda}_g}$ and $\dot{\Lambda}_g \equaldef \sum_{i=g}^N \lambda_i^{+_1, \cdots, +_{g-1}} - \sum_{i=G+1}^N \lambda_i^{+_1, \cdots, +_G}$. \medskip

Introducing $\dot{\Lambda}_{G+1}=0$ and then $\Lambda_g \equaldef \dot{\Lambda}_{g+1} + \sum_{i=G+1}^N \lambda_i^{+_1, \cdots, +_G} = \sum_{i=g+1}^N \lambda_i^{+_1, \cdots, +_{g}}\geq 0$ (with equality only if $g=G=N$), the probability simplifies to\footnote{As in the example with 2 individuals, it can be checked from the integral computation that the limit for a term in a denominator converging to $0$ gives the correct probability at the points where the function above is undefined. These cases correspond to knife-edge cases in which a move does not change the rate at which the next event occurs: enhanced rates due to peer effects exactly offset the withdrawal from the current mover.}

\begin{equation}
    \prod_{i=1}^G \lambda_i^{+_1, \cdots, +_{i-1}} \sum_{g=0}^{G} \frac{e^{-\Lambda_g S}}{\prod_{h \neq g} \Lambda_h - \Lambda_g}
\end{equation}
\end{proof}

\subsection*{Proof of Lemma \ref{identification_lemma}}

\begin{proof}

The proof considers the two cases separately. 

\paragraph{Identification under block structure}

In what follows, $\vec{0}$ denotes a $0$-vector of comformable size. Consider the identified probability $\mathds{P}[Y = \vec{0}\vert X=x, W] = e^{-\sum_{i=1}^n (\lambda_i(x_i))S}$. Setting $x_i=x \ \forall i$, $\lambda(x) = \frac{-\ln(p_{00}(x, \ldots, x))}{n S}$ and thus the baseline rates are identified. 

Then, $p_{10}(x_1, x_2) = \lambda_1 e^{-\lambda_2^+ S} g(\lambda_1+\lambda_2-\lambda_2^+)$ allows us to identify $\lambda_2^+$ nonparametrically as $\lambda^+(x_2, x_1)$ because $e^{-x S} g(\lambda_1+\lambda_2-x) = e^{-(\lambda_1+\lambda_2) S} g(-(\lambda_1+\lambda_2-x))$ is an invertible function of $x$ for any $\lambda_1, \lambda_2$. 

Consider now
\begin{align*}
\begin{split}
    \mathds{P}[y_1 = 1, y_{-1}=0\vert X, W] = &\lambda_1 \left(\frac{e^{-\sum_{i=2}^N \lambda_i^{+{1}} S}}{\sum_{i=1}^N \lambda_i - \sum_{i=2}^N \lambda_i^{+{1}}} - \frac{e^{-\sum_{i=1}^N \lambda_i S}}{\sum_{i=1}^N \lambda_i - \sum_{i=2}^N \lambda_i^{+{1}}}  \right) \\ 
    & = \lambda_1 e^{-\sumin \lambda_i S} g\left(\sum_{i=2}^N \lambda_i^{+{1}} - \sumin \lambda_i\right)
\end{split}
\end{align*}
where $g$ is defined as in Example \ref{two_people}. 

Setting $x_j = x_i$ for all $j \neq 1$, $\mathds{P}[y_1 = 1, y_{-1}=0\vert X=(x_1, x, \ldots, x), W] = \lambda_1 e^{-\sumin \lambda_i S} g\left( \sum_{j: W_{1j}=1} \lambda_i^{+{1}} - \sum_{j: W_{1j}=1}\lambda_i \right) = \lambda_1 e^{-\sumin \lambda_i S} g\left( \vert {j: W_{1j}=1}\vert \lambda_i^{+{1}} - \vert {j: W_{1j}=1}\vert \lambda_i \right)$ and thus $\lambda_i^{+{1}}$ is identified by invertibility of $g(\cdot)$ and identification of $\lambda_i$. 

Proceeding iteratively, $\lambda_i^{+\mathcal{F}}$ is identified from the conditional probabilities and the knowledge of all $\lambda_i^{+{F}}$ with $\vert F\vert < \vert \mathcal{F} \vert$.

\paragraph{Identification under sparsity}

The proof shows how to recover the rates from the identified objects. Covariates are handled by conditioning throughout and setting peers to similar values, as in the proof under block structure. The proof again proceeds iteratively by showing how to identify baseline rates, then (once-) updated rates under the knowledge of baseline rates, etc. 

Note that we observe at least $K$ adopters for any $K \in \mathds{R}$ with probability approaching one. Because degrees are bounded, there is a finite number of $\lambda^{+\mathcal{F}}$ so that each rate can be mapped into the set $\{1, \ldots, n_\lambda\}$ by some function $m$. 


Let the index of the rate of the $k$-th adopter (at their adoption time) be denoted by $\Lambda_k$. For any $K \in \mathds{N}$, $\lambda$ is given by the inverse of $f_K(\lambda) = \frac{1}{K} \sum_{k=1}^K \mathds{P}(\Lambda_k=m(\lambda))$, which is strictly increasing in $\lambda$ if the network is connected.  
This function is identified from $\frac{1}{K} \sum_{k=1}^K \mathds{1}(\Lambda_k=r(\lambda))-\mathds{P}(\Lambda_k=r(\lambda)) \rightarrow^p 0$ by the law of large numbers, noting the independence and integrability. 





\end{proof}

\subsection*{Proof of Theorem \ref{consistency_theorem} and \ref{normality_theorem}}

I provide divide the proof into two sections depending on the type of asymptotics: several networks and one large network. 

\subsubsection*{Asymptotics with several networks}

\begin{proof}

The proofs verify \citet{newey1994large}'s sufficient conditions for consistency and asymptotic normality of extremum estimators. \medskip


The parameter space is compact by assumption. The log-likelihood of block $b$ is $l_b = \sum_{y \in \{0,1\}^{N_b}} \mathds{1}[Y_b=y] \ln(\mathds{P}[Y_b=y])$, where the elements of the sum are bounded from below using that $N_b \leq \overline{N}$ and compactness of the rates and from above by 0. Since in addition $\mathds{P}[Y_b=y]$ is continuous, uniform laws of large numbers implies (i) $\sup_{\{\lambda\}} \vert B^{-1} \sum_b l_b - \mathds{E}[l_b] \vert \rightarrow 0$ and (ii) $\mathds{E}[l_b]$ is continuous. 

By continuity of the Hessian matrix and compactness of the parameter space, it follows that 
\begin{equation}
    \sup_{\{\lambda\}} \vert H_n - H \vert \rightarrow 0
\end{equation}
where $H$ is continuous. Asymptotic normality of the score follows from laws of large numbers for triangular arrays, noting that $\mathds{V}[l_b] \rightarrow \sigma^2$

\bigskip

By assumption, the parameters are identified and live in a compact set. The objective function converges uniformly to $\mathds{E}[l_b]$, which is continuous, by uniform law of large numbers (\textit{e.g.}, Lemma 2.4 in \citet{newey1994large}). Indeed, $l_b$ is continuous and 

\begin{align}
    \begin{split}
        \min_{y, p, \lambda} \mathds{P}[p(\lambda)] \leq e^{l_b} \leq 1, 
    \end{split}
\end{align}
where the minimization is over a finite set for $y, p$ since $N_b \leq \overline{N}$ and the minimization over $\lambda$ is over a compact set for a continuous function, so that $\sup \vert B^{-1} \sum_{b=1}^B l_b - \mathds{E}[l_b]\vert \rightarrow 0$. 

\begin{equation}
    \frac{1}{\sqrt{B}} \sum_{b=1}^B s_b \rightarrow^d \mathcal{N}(0; \mathds{E}[s_b s_b'])
\end{equation}
noting that the score has mean zero and 
\begin{align}
    \begin{split}
        \mathds{V}[s_b] & = \mathds{E}[s_b s_b'] \\
        & \leq C \mathds{E}[\partial L_b \partial L_b'] \\
        & = C \sum_{W_b} \mathds{P}[W_b] \mathds{E}[\partial L_b \partial L_b'\vert W_b] \\
        & = C \sum_{W_b} \mathds{P}[W_b] \sum_{\tilde{y}} \mathds{P}[Y = \tilde{y}] \partial L_b(\tilde{y}) \partial L_b'(\tilde{y}) \\ 
        & \leq C \max_{\tilde{y}} \partial L_b(\tilde{y}) \partial L_b'(\tilde{y})
    \end{split}
\end{align}

\end{proof}

\pagebreak

\subsubsection*{Asymptotics with a large network}

\begin{proof}

The likelihood of the ordering $(1), \ldots, (G)$ is given by
\begin{equation}
    \prod_{i=1}^G \lambda_i^{+(1), \cdots, +(i-1)} \sum_{g=0}^{G} \frac{e^{-\Lambda_g S}}{\prod_{h \neq g} \Lambda_h - \Lambda_g} = \sum_{g=0}^{G} \frac{e^{-R_g}}{\prod_{h \neq g} R_h - R_g}
\end{equation}
where $R_g \equaldef \Lambda_g S - \sum_{i=1}^G \ln(\lambda_i^{+(1), \cdots, +(i-1)} S)$

The sum can be recognized as the $G$-th divided difference of the exponential function $e^{-x}$, multiplied by $(-1)^G$. By the Hermite-Genocchi formula, noting that the $G$-th derivative of $e^{-x}$ is $(-1)^G e^{-x}$, we obtain
\begin{equation}
    \sum_{g=0}^{G} \frac{e^{-R_g S}}{\prod_{h \neq g} R_h - R_g} = \int_{\Delta_G} \mbox{exp}\left(-\sum_{g=0}^G t_g R_g\right) \mbox{d}t
\end{equation}
where $\Delta_G$ is the $G$-dimensional simplex. 

Define the integral on the scaled simplex as a function $F$:
\begin{equation}
    F(u) \equaldef \int_{t_g \geq 0, \sum_{g=0}^G t_g = u} \mbox{exp}\left(-\sum_{g=0}^G t_g R_g\right) \mbox{d}t
\end{equation}

The Laplace transform of $F$ is available in closed form:
\begin{equation}
    \mathcal{L}F(s) = \prod_{g=0}^G \frac{1}{R_g+s}
\end{equation}

As a result, the likelihood, which coincides with $F(1)$, can be obtained from an inverse Laplace transform, and the convergence can be interpreted as a saddle point problem:
\begin{align}
\begin{split}
    \int_{\Delta_G} \mbox{exp}\left(-\sum_{g=0}^G t_i R_g\right) \mbox{d}t & = F(1) \\
    & = \frac{1}{2 \pi i} \int_{c-i \infty}^{c+i \infty} \frac{e^s}{\prod_{g=0}^G (R_g + s)} \mbox{d}s \\
    & = \frac{1}{2 \pi i N^{G-1}} \int_{c_0-i \infty}^{c_0+i \infty} \frac{(e^u)^N}{\prod_{g=0}^G (R_g/N + u)} \mbox{d}u \\
    & = \frac{1}{2 \pi i N^{G-1}} \int_{c_0-i \infty}^{c_0+i \infty} \left(e^{z - \frac{1}{N} \sum_{g=0}^G \ln(R_g/N + z)}\right)^N \mbox{d}z
\end{split}
\end{align}

Deforming the contour of integration to obtain the real saddle point and applying the Laplace principle, it is seen that the integral concentrates around the $z$ that satisfies the stationarity condition $\meangG \left(R_g/N) + z\right)^{-1} = 1$.

This implies that
\begin{align}\label{likelihood_approx}
\begin{split}
    l_n \equaldef N^{-1} \ln\left(G! \vert \mathcal{P}\vert^{-1} \sum_{p \in \mathcal{P}} L_p\right) & = z_N^{*} - \meangG \ln\left(R_g/N + z_N^{*}\right) \\
    & + \frac{G}{N} \ln\left( \frac{G}{N} \right) - \frac{G}{N} \\
    & + O(N^{-1} \ln(N)) + O(G^{-1} \ln(G))
\end{split}
\end{align}
where $z_N^{*}$ solves $\meangG \left(\hat{r}(g/N) + z_N^{*}\right)^{-1} = 1$.

The score is given by
\begin{equation}
    s_N \equaldef - \meangG \frac{\partial_\theta R_g/N}{R_g/N + z_N^{*}} + O(N^{-1} \ln(N))
\end{equation}
while the Hessian is
\begin{equation}
    H_N \equaldef - \meangG \left( \frac{\partial_{\theta \theta'} R_g/N}{R_g/N + z_N^{*}} - \frac{\partial_{\theta} R_g/N \partial_{\theta'} R_g/N}{(R_g/N + z_N^{*})^2} - \frac{\partial_{\theta} R_g/N \partial_{\theta'} z_N(\theta)}{(R_g/N + z_N^{*})^2}\right)+ O(N^{-1} \ln(N))
\end{equation}

Applying the implicit function theorem on the condition $\meangG \left(R_g/N + z_N^{*}\right)^{-1} = 1$ delivers $\partial_{\theta'} z_N^*(\theta) = - \left(\meangG \frac{1}{(R_g/N + z_N^{*})^2}\right)^{-1} \meangG \frac{\partial_{\theta'} R_g/N}{(R_g/N + z_N^{*})^2}$.

The proof proceeds by a classical expansion around the true value of $\theta$, we have 
\begin{align}
\begin{split}
        \sqrt{N} (\hat{\theta}-\theta) & = \left(H_N\right)^{-1} \sqrt{N} s_N \\
        & \overset{d}{\longrightarrow} \mathcal{N}(0, H^{-1})
\end{split}
\end{align}
since, as argued below, $\sup_\theta \vert H_n - H\vert \overset{p}{\longrightarrow} 0$ and the score converges in distribution to a normal random variable.

Let 
\begin{equation}
    \hat{r}(g/N) \equaldef \frac{R_g}{N}
\end{equation}
and
\begin{equation}
    \rho(g/N) \equaldef \sqrt(N) (\partial_\theta R_g/N - \mathds{E}[\partial_\theta R_g/N])
\end{equation}
and the functions are extended in the usual way to the unit interval by interpolating.

The variable $R_g$ can be decomposed as
\begin{equation}\label{R_g_decomposition}
    R_g = \sum_{h=1}^g \Delta_h = \sum_{h=1}^g \Delta_h - \mathds{E}[\Delta_h\vert \mathcal{F}_{h-1}] + \sum_{h=1}^g \mathds{E}[\Delta_h\vert \mathcal{F}_{h-1}] - \mathds{E}[\Delta_h] + \mathds{E}[R_g]
\end{equation}
with $\Delta_h \equaldef R_h-R_{h-1}$, noting that $\mathds{E}[R_0]=R_0$

Since the terms in each sum are bounded and uncorrelated, the law of large numbers implies convergence of $R_g/N$ to its expectation. Furthermore, the convergence is uniform in $\theta$ since the terms are continuous in $\theta$ and uniformly bounded by $2 \sup_{i, \mathcal{F}} \left\vert - \lambda_{(i)}^{\mathcal{F}} + \sum_j \lambda_{(j)}^{\mathcal{F} \cup \{i\}} - \lambda_{(j)}^{\mathcal{F}} \right\vert \leq ((\overline{d}+1)\overline{\lambda} - \overline{d} \underline{\lambda})$, where $\overline{\lambda} = \sup_{i, \mathcal{F}} \lambda_i^{\mathcal{F}}$ and $\underline{\lambda} = \inf_{i, \mathcal{F}} \lambda_i^{\mathcal{F}}$. As a result,

\begin{equation}
    \sup_\theta \vert \hat{r}(z) - r(z)\vert \overset{p}{\longrightarrow} 0,
\end{equation}
where $r(z) = \lim_{N \rightarrow \infty} \mathds{E}[R_{z*N}g/N]$.

In addition, 
\begin{equation}
    \sup_\theta \vert z_N^* - z^*\vert \overset{p}{\longrightarrow} 0
\end{equation}
because $z_N^*$ is the maximizer of a strictly concave function which converges uniformly to the function that $z^*$ maximizes.

Then, as $G/N \overset{p}{\rightarrow} c \in ]0, 1[$, 
\begin{equation}
    l_N \overset{p}{\longrightarrow} l \equaldef z^* - \int_0^c \ln(r(z)+z^*) \mbox{d}z + c \ln(c) - c
\end{equation}


The score satisfies 
\begin{align}
    \begin{split}
        \sqrt{N} \meangG \frac{\partial \hat{r}}{\hat{r}+z_N^*} = \sqrt{N} \int_{0}^{G/N} \frac{\partial \hat{r}}{\hat{r}+z_N^*} + o_p(1)
    \end{split}
\end{align}
and since $\vert \hat{r}+z_N^*-(r+z^*)\vert \leq \vert \hat{r}-r\vert + \vert z_N^*-z^*\vert \overset{p}{\longrightarrow} 0$, where the convergence for $\hat{r}$ is uniform over $z \in [0, 1]$, this becomes
\begin{align}
    \begin{split}
        \sqrt{N} \meangG \frac{\partial \hat{r}}{\hat{r}+z_N^*} & = \sqrt{N} \int_{0}^{G/N} \frac{\partial \hat{r}(z)}{\hat{r}(z)+z_N^*} + o_p(1) \\
        & = \sqrt{N} \int_{0}^{G/N} \frac{\partial \hat{r}(z)}{r(z)+z^*} \mbox{d}z+ o_p(1) \\ 
        & = \sqrt{N} \int_{0}^{c} \frac{\partial \hat{r}(z)}{r(z)+z^*} \mbox{d}z+ \sqrt{N} \int_{c}^{G/N} \frac{\partial \hat{r}(z)}{r(z)+z^*} \mbox{d}z + o_p(1) \\ 
        & = \sqrt{N} \int_{0}^{c} \frac{\partial \hat{r}(z)}{r(z)+z^*} \mbox{d}z + o_p(1) \\
    \end{split}
\end{align}
noting that $\left\vert \sqrt{N} \int_{c}^{G/N} \frac{\partial \hat{r}(z)}{r(z)+z^*} \mbox{d}z \right\vert \leq \vert G/N - c \vert \sqrt{N} \sup_{z \in [0, 1]} \frac{\partial \hat{r}(z)}{r(z)+z^*}$.

As a result, if $\rho$ converges to a process $\rho^*$, the continuous mapping theorem implies
\begin{equation}
    \sqrt{N} s_N = \int_0^c \frac{\rho(z)}{r(z)+z^*} \mbox{d}z + o_P(1) \overset{d}{\longrightarrow} \int_{0}^c \omega(z) \rho^*(z) \mbox{d}z
\end{equation}
where $\omega(z) \equaldef (r(z)+z^*)^{-1}$ is a function function such that $\int_0^c \omega(z) = 1$ by the stationarity condition.

It remains to establish convergence of $\rho$ to a process and examining its structure. Decomposing in a similar way as for $R_g$ in (\ref{R_g_decomposition}), $\rho(z)$ is equal to $ \sum_{h=1}^{\floor*{Nz}} \Delta_h^\rho$ up to negligible terms. 

Tightness for $\rho$ can be verified by noting (i) that the increments are bounded and thus jumps are vanishing, $\sup_{h} \vert \Delta_h^\rho \vert/N^{1/2}\leq C/N^{1/2} \overset{p}{\longrightarrow} 0$, and (ii) convergence of the variation $\frac{1}{N} \sum_{h=1}^{\floor{N z}} \mathds{E}[(\left(\Delta_h^\rho\right)^2 \vert \mathcal{F}_{h-1}]$:
\begin{equation}
    \sup_{z \in [0, 1]} \left\vert \frac{1}{N} \sum_{h=1}^{\floor{N z}} \mathds{E}[(\left(\Delta_h^\rho\right)^2 \vert \mathcal{F}_{h-1}] - \sigma^2(z) \right\vert \overset{p}{\longrightarrow} 0
\end{equation}

This establishes Aldous' criterion and hence tightness \citep{whitt2007proofs}. In addition, a standard argument using a central limit theorem for (bounded) martingale difference sequences implies that the finite-dimensional distributions converge to a normal distribution. This establishes the convergence of $\rho$ to a process $\rho^*$ in $[0, 1]$. The stochastic process $\rho^*$ is $B(\sigma^2(\cdot))$, in particular it has mean 0, continuous paths, and covariance structure $\sigma^2(t \wedge s)$. 

Finally, using the uniform convergence results on $\hat{r}$ and $z_N^*$, we have
\begin{align}
\begin{split}
    \sup_{\theta} \vert H_N - H \vert \overset{p}{\longrightarrow} 0 
\end{split}
\end{align}
where
\begin{equation}
    H \equaldef - \int_0^c \frac{\partial_{\theta \theta'} r(z)}{r(z)+z^*} \mbox{d}z + \int_0^c \frac{\partial_{\theta} r(z) \partial_{\theta'} r(z)}{(r(z)+z^*)^2}  \mbox{d}z - \frac{\int_0^c \frac{\partial_{\theta} r(z)}{(r(z)+z^*)^2} \mbox{d}z \ \int_0^c \frac{\partial_{\theta'} r(z)}{(r(z)+z^*)^2} \mbox{d}z}{\int_0^c \frac{1}{(r(z)+z^*)^2} \mbox{d}z}.
\end{equation}

\end{proof}

\pagebreak

\section*{Appendix B: Simulations}

I now assess the performance of the estimator in simulations. I first consider correctly specified models. In the second subsection, I investigate the robustness of the estimator to omitted variables, measurement errors, and group heterogeneities. 

\subsection*{Simulations with block and homophilic network formation}

I simulate the stochastic process described in Section 2.2 with an underlying network structure of either 'classrooms' or homophilic matching type, both of which are common in empirical studies. \medskip

First, I construct a network with 1000 individuals and a 'block' structure ($(W=I \otimes \iota \iota')$) with groups of size 5, 10, and 20. In the previous section's notation, this means $N=1000$, $n_b=5 \ \forall b$, and $B = 1000/n_b$ and individuals are connected to all individuals within the same block. 
I set the family of rates to obey (\ref{parametric_spec_for_lambdas}) with three covariates (an intercept, a uniform variable on $[-1;1]$ and a (standard) normal variable, respectively), various levels of peer effect strength ($\delta \in \{-0.5, 0, 0.5\}$), and $\beta = \begin{pmatrix} -1 \\ 1 \\ 0.5 \end{pmatrix}$. \medskip

There is no information about the order of adoptions so all permutations \textit{a priori} matter. I make use of the random sampling over permutations mentioned in Section 2 to alleviate the computational burden whenever the number of adopted people in a group exceeds 8. \medskip

Estimates are compared to SAR estimates from a simple (endogenous) regression on $x_i$ and $W_i y$ and to the SAR maximum likelihood estimator\footnote{The weighting matrix is row-normalized since the process suggests peer effects depend on the average number of adopted friends. Notice, however, that the model using sums has an equivalent representation using averages when groups have the same size: it amounts to scaling $\delta$ by group size.}
\medskip

The results are reported in Table 1. The maximum of likelihood estimator described in the previous section performs well in all instances and exhibits very low bias. A standard regression is usually able to pick up the correct sign of peer effects in this specific setup, but cannot recover the structural coefficient. The SAR MLE broadly follows the same lines.

\begin{table}[!ht]\caption{Simulations with 'classrooms' network structure}
\[\begin{array}{ | l | l | l | l | l | l | l | l | l | l | l | l | }
\hline
 & & & \multicolumn{3}{c}{\mbox{Bias}} & \multicolumn{3}{c}{\mbox{Standard deviation}} & \multicolumn{3}{c}{\mbox{RMSE}} \\ \hline
n_b & \delta & & \mbox{Reg} & \mbox{SAR} & \mbox{Exp} & \mbox{Reg} & \mbox{SAR} & \mbox{Exp} & \mbox{Reg} & \mbox{SAR} & \mbox{Exp} \\ \hline 
\multirow{9}{*}{5} & \multirow{3}{*}{-0.5} & \hat{\delta} & 0.39 & 0.43 & 0.03 & 0.04 & 0.03 & 0.12 & 0.39 & 0.43 & 0.13\\ \cline{3-12}
& & \hat{\beta}_1 & -0.70 & -0.37 & 0.00 & 0.02 & 0.03 & 0.09 & 0.70 & 0.38 & 0.09\\ \cline{3-12}
& & \hat{\beta}_2 & -0.36 & -0.19 & 0.00 & 0.01 & 0.02 & 0.05 & 0.36 & 0.20 & 0.05\\ \cline{2-12}
& \multirow{3}{*}{0} &\hat{\delta} & -0.02 & -0.01 & 0.00 & 0.08 & 0.04 & 0.10 & 0.08 & 0.04 & 0.10\\ \cline{3-12}
& & \hat{\beta}_1 & -0.70 & -0.37 & 0.00 & 0.02 & 0.03 & 0.09 & 0.70 & 0.37 & 0.09\\ \cline{3-12}
& & \hat{\beta}_2 & -0.36 & -0.20 & 0.00 & 0.01 & 0.02 & 0.05 & 0.36 & 0.20 & 0.05\\ \cline{2-12}
& \multirow{3}{*}{0.5} &\hat{\delta} & -0.35 & -0.41 & -0.01 & 0.07 & 0.04 & 0.09 & 0.36 & 0.42 & 0.09\\ \cline{3-12}
& & \hat{\beta}_1 & -0.71 & -0.37 & 0.00 & 0.02 & 0.03 & 0.09 & 0.71 & 0.37 & 0.09\\ \cline{3-12}
& & \hat{\beta}_2 & -0.36 & -0.21 & 0.00 & 0.01 & 0.02 & 0.05 & 0.36 & 0.21 & 0.05\\ \cline{1-12}
\multirow{9}{*}{10} & \multirow{3}{*}{-0.5} & \hat{\delta} & 0.32 & 0.41 & 0.00 & 0.14 & 0.07 & 0.13 & 0.35 & 0.41 & 0.13\\ \cline{3-12}
& & \hat{\beta}_1 & -0.69 & -0.37 & 0.00 & 0.02 & 0.04 & 0.09 & 0.69 & 0.38 & 0.09\\ \cline{3-12}
& & \hat{\beta}_2 & -0.35 & -0.19 & 0.00 & 0.01 & 0.02 & 0.05 & 0.35 & 0.20 & 0.05\\ \cline{2-12}
& \multirow{3}{*}{0} &\hat{\delta} & -0.01 & -0.01 & 0.00 & 0.12 & 0.07 & 0.11 & 0.12 & 0.07 & 0.11\\ \cline{3-12}
& & \hat{\beta}_1 & -0.70 & -0.37 & 0.00 & 0.02 & 0.04 & 0.09 & 0.70 & 0.37 & 0.09\\ \cline{3-12}
& & \hat{\beta}_2 & -0.36 & -0.20 & 0.00 & 0.01 & 0.02 & 0.05 & 0.36 & 0.20 & 0.05\\ \cline{2-12}
& \multirow{3}{*}{0.5} &\hat{\delta} & -0.37 & -0.42 & 0.00 & 0.10 & 0.06 & 0.10 & 0.38 & 0.43 & 0.10\\ \cline{3-12}
& & \hat{\beta}_1 & -0.71 & -0.37 & 0.01 & 0.02 & 0.04 & 0.08 & 0.71 & 0.37 & 0.08\\ \cline{3-12}
& & \hat{\beta}_2 & -0.36 & -0.21 & 0.01 & 0.01 & 0.02 & 0.06 & 0.36 & 0.21 & 0.06\\ \cline{1-12}
\multirow{9}{*}{20} & \multirow{3}{*}{-0.5} & \hat{\delta} & 0.30 & 0.40 & 0.00 & 0.21 & 0.10 & 0.13 & 0.36 & 0.41 & 0.13\\ \cline{3-12}
& & \hat{\beta}_1 & -0.70 & -0.37 & 0.01 & 0.02 & 0.06 & 0.08 & 0.70 & 0.37 & 0.09\\ \cline{3-12}
& & \hat{\beta}_2 & -0.36 & -0.20 & 0.01 & 0.01 & 0.02 & 0.05 & 0.36 & 0.20 & 0.05\\ \cline{2-12}
& \multirow{3}{*}{0} &\hat{\delta} & -0.06 & -0.03 & -0.01 & 0.18 & 0.10 & 0.11 & 0.19 & 0.10 & 0.11\\ \cline{3-12}
& & \hat{\beta}_1 & -0.70 & -0.36 & 0.00 & 0.02 & 0.06 & 0.08 & 0.70 & 0.36 & 0.08\\ \cline{3-12}
& & \hat{\beta}_2 & -0.36 & -0.20 & 0.00 & 0.01 & 0.02 & 0.05 & 0.36 & 0.20 & 0.05\\ \cline{2-12}
& \multirow{3}{*}{0.5} &\hat{\delta} & -0.38 & -0.42 & -0.02 & 0.15 & 0.09 & 0.10 & 0.40 & 0.43 & 0.11\\ \cline{3-12}
& & \hat{\beta}_1 & -0.71 & -0.36 & 0.07 & 0.02 & 0.06 & 0.18 & 0.71 & 0.37 & 0.19\\ \cline{3-12}
& & \hat{\beta}_2 & -0.36 & -0.21 & 0.03 & 0.01 & 0.02 & 0.08 & 0.36 & 0.21 & 0.09\\ \hline
\end{array}\]
\end{table}

\pagebreak 

I now consider another network structure, in which individuals within groups decide whether to make a connection based on their characteristics. Specifically, I consider a homophilic link formation process in which individual match according to their similarities: $W_{ij}=1$ iff $\frac{\Vert X_{1i} - X_{1j}\Vert + \Vert X_{2i} - X_{2j}\Vert}{2} < \eta_{ij})$, where the collection of $\eta_{ij}=\eta_{ji}$ forms an array of independent uniform random variables. Group sizes are $5, 20,$ or a larger group of $100$ and the sample size is gain $N=1000$. \medskip

The results are displayed in the next table.

\begin{table}[!ht]\caption{Homophilic}
\[\begin{array}{ | l | l | l | l | l | l | l | l | l | l | l | l | }
\hline
 & & & \multicolumn{3}{c}{\mbox{Bias}} & \multicolumn{3}{c}{\mbox{Standard deviation}} & \multicolumn{3}{c}{\mbox{RMSE}} \vline \\ \hline
n_b & \delta & & \mbox{Reg} & \mbox{SAR} & \mbox{Exp} & \mbox{Reg} & \mbox{SAR} & \mbox{Exp} & \mbox{Reg} & \mbox{SAR} & \mbox{Exp} \\ \hline 
\multirow{9}{*}{5} & \multirow{3}{*}{-0.5} & \hat{\delta} & 0.39 & 0.42 & 0.03 & 0.04 & 0.03 & 0.13 & 0.39 & 0.43 & 0.13\\ \cline{3-12}
& & \hat{\beta}_1 & -0.70 & -0.38 & -0.01 & 0.02 & 0.02 & 0.09 & 0.70 & 0.38 & 0.09\\ \cline{3-12}
& & \hat{\beta}_2 & -0.36 & -0.20 & 0.00 & 0.01 & 0.02 & 0.05 & 0.36 & 0.20 & 0.05\\ \cline{2-12}
& \multirow{3}{*}{0} &\hat{\delta} & 0.00 & 0.00 & -0.02 & 0.04 & 0.03 & 0.11 & 0.04 & 0.03 & 0.11\\ \cline{3-12}
& & \hat{\beta}_1 & -0.69 & -0.38 & 0.02 & 0.02 & 0.02 & 0.09 & 0.69 & 0.38 & 0.09\\ \cline{3-12}
& & \hat{\beta}_2 & -0.36 & -0.19 & 0.01 & 0.01 & 0.02 & 0.05 & 0.36 & 0.20 & 0.05\\ \cline{2-12}
& \multirow{3}{*}{0.5} &\hat{\delta} & -0.38 & -0.41 & -0.03 & 0.03 & 0.02 & 0.10 & 0.38 & 0.41 & 0.11\\ \cline{3-12}
& & \hat{\beta}_1 & -0.71 & -0.38 & 0.02 & 0.02 & 0.02 & 0.08 & 0.71 & 0.38 & 0.08\\ \cline{3-12}
& & \hat{\beta}_2 & -0.36 & -0.20 & 0.00 & 0.01 & 0.02 & 0.05 & 0.36 & 0.21 & 0.05\\ \cline{1-12}
\multirow{9}{*}{10} & \multirow{3}{*}{-0.5} & \hat{\delta} & 0.41 & 0.44 & 0.03 & 0.05 & 0.03 & 0.12 & 0.41 & 0.44 & 0.12\\ \cline{3-12}
& & \hat{\beta}_1 & -0.69 & -0.39 & 0.00 & 0.02 & 0.02 & 0.09 & 0.70 & 0.39 & 0.09\\ \cline{3-12}
& & \hat{\beta}_2 & -0.35 & -0.20 & 0.00 & 0.01 & 0.02 & 0.05 & 0.35 & 0.20 & 0.05\\ \cline{2-12}
& \multirow{3}{*}{0} &\hat{\delta} & 0.00 & 0.00 & -0.02 & 0.04 & 0.03 & 0.10 & 0.04 & 0.03 & 0.10\\ \cline{3-12}
& & \hat{\beta}_1 & -0.70 & -0.38 & 0.00 & 0.02 & 0.02 & 0.08 & 0.70 & 0.38 & 0.08\\ \cline{3-12}
& & \hat{\beta}_2 & -0.36 & -0.20 & 0.00 & 0.01 & 0.02 & 0.05 & 0.36 & 0.20 & 0.05\\ \cline{2-12}
& \multirow{3}{*}{0.5} &\hat{\delta} & -0.39 & -0.42 & -0.05 & 0.05 & 0.03 & 0.10 & 0.39 & 0.43 & 0.11\\ \cline{3-12}
& & \hat{\beta}_1 & -0.71 & -0.37 & 0.01 & 0.02 & 0.02 & 0.08 & 0.71 & 0.37 & 0.08\\ \cline{3-12}
& & \hat{\beta}_2 & -0.36 & -0.21 & 0.00 & 0.01 & 0.02 & 0.05 & 0.36 & 0.21 & 0.05\\ \hline
\end{array}\]
\end{table}

Although the performance of OLS or SAR-MLE in terms of bias and RMSE in the absence of peer effects ($\delta=0$) suggests that these estimators may successfully detect the absence of social influence, notice that estimates are generally attenuated compared to the structural parameter and that decisions will eventually be based on tests or confidence intervals. As a result, the coverage performance of the confidence intervals may be a more relevant benchmark and will be analyzed in the next subsection. 

Interestingly, OLS and SAR-MLE feature attenuation bias with respect to the structural parameter. As a result, they may seem to perform better in terms of RMSE in the absence of peer effect.
In practice, however, what matters is the test for the presence of peer effect or, equivalently, the resulting confidence intervals. In the next subsection, I explore the coverage performance of the three estimators to assess their ability to (correctly) not reject a null hypothesis of no peer effects in both correctly specified and misspecified models.

\subsection*{Mispecifications type of results}

Peer effect studies are often subject to criticism due to modeling \citep{manski1993identification, angrist2014perils} and empirical \citep{angrist2014perils} concerns. While it is hoped that the framework developed in this paper alleviates modeling concerns - in particular, by avoiding reflection problems -, it is of interest to evaluate the behavior of the estimator under frequent empirical difficulties: missing or omitted covariates, group level heterogeneity, or measurement error. \medskip

I focus here on the peer effect parameter $\delta$, which will typically the parameter of interest. \medskip

Because OLS and SAR cannot identify the structural coefficient but could still detect the presence of peer effects, it is of interest to look at the coverage performance. I look at the frequency at which a $95\%$ confidence interval contains $0$, indicating the absence of peer effects, under the generating process in which peer effects are indeed absent ($\delta=0)$. \medskip

Table \ref{mispecification_table} reports the coverage of a $95\%$ confidence interval under the homophilic network structure when the researcher (i) observes both covariates, (ii) observes only the first covariate, (iii) observes a mismeasured (with ($\mathcal{N}(0;0.25)$) error) first covariate, and (iv)/(v)/(vi) there is (uniform on $[-1;0]$) group heterogeneity (added to the argument of the exponential) in the (i)/(ii)/(iii) scenario.  \medskip

\begin{table}[!ht]\caption{Coverage analysis with potential misspecification}\label{mispecification_table}
\[\begin{array}{ | l | l | l | l | l | l |}
\hline
 & & & \multicolumn{3}{c}{\mbox{Coverage}} \vline \\ \hline
n_b & \delta & & \mbox{Reg} & \mbox{SAR} & \mbox{Exp} \\ \hline 
\multirow{6}{*}{5} & \multirow{6}{*}{0} & \mbox{Size} & 0.90 & 0.79 & 0.95\\ \cline{3-6}
& & \mbox{Size} & 0.72 & 0.62 & 0.90\\ \cline{3-6}
& & \mbox{Size} & 0.69 & 0.55 & 0.90\\ \cline{3-6}
& & \mbox{Size} & 0.83 & 0.70 & 0.89\\ \cline{3-6}
& & \mbox{Size} & 0.62 & 0.47 & 0.71\\ \cline{3-6}
& & \mbox{Size} & 0.56 & 0.42 & 0.67\\ \hline
\end{array}\]
\caption{Coverage performance of a 95\% confidence interval from OLS with clustered standard errors, SAR-MLE, and maximum of likelihood on latent exponential processes.}
\end{table}

The coverage performance of the estimator developed in the paper is far better than that of OLS and SAR-MLE. Although the most serious issues (lack of covariate and measurement error combined with heterogeneity issues) can lead to severe size distortions, spurious peer effects are unlikely under more standard scenarios. The test for the presence of peer effect is adequately sized in the case of correct specification and is moderately distorted under measurement error or group heterogeneity. \medskip 

The inability to control for any covariate is the most problematic issue as it is exacerbated by homophilic matching, in which case peer effects are spuriously ascribed through correlation of outcomes among similar groups. \medskip

Both OLS and SAR-MLE have a tendency to spuriously detect peer effects at a rate higher than the pre-specified level, even with homogeneous groups and adequate covariates. Any empirical difficulty such as measurement error, unobserved covariate, or heterogeneity leads by itself to a high risk of unwarranted rejection of the null of no peer effects, echoing critiques in \citet{angrist2014perils}.

\end{document}